# Modular Landfill Remediation for AI Grid Resilience

Qi He
Technical Infrastructure
Google
USA
qhe11@fordham.edu

Chunyu Qu
Public Sector
Dun & Bradstreet Inc.
USA
quc@dnb.com

*Abstract*— Rising AI electricity demand and persistent landfill methane emissions constitute coupled constraints on U.S. digital infrastructure and decarbonization. While China has achieved a rapid "de-landfilling" transition through centralized coordination, the U.S. remains structurally "locked in" to landfilling due to fragmented governance and carbon accounting incentives. This paper proposes a modular legacy landfill remediation framework to address these dual challenges within U.S. institutional constraints. By treating legacy sites as stock resources, the proposed system integrates excavation, screening, and behind-the-meter combined heat and power (CHP) to transform environmental liabilities into resilience assets. A system analysis of a representative AI corridor demonstrates that such modules can mitigate site-level methane by 60-70% and recover urban land, while supplying approximately 20 MW of firm, islandable power. Although contributing only approximately 5% of a hyperscale data center's bulk load, it provides critical microgrid resilience and black-start capability. We conclude that remediation-oriented waste-to-energy should be valued not as a substitute for bulk renewables, but as a strategic control volume for buffering critical loads against grid volatility while resolving long-term environmental liabilities.

## I. Introduction

Rapid growth in urban populations, production, and consumption is driving an unprecedented expansion of municipal solid waste (MSW). The World Bank projects global MSW to increase from about 2.01 billion tonnes in 2016 to 3.40 billion tonnes by 2050, with roughly one-third of current waste still mismanaged[1]. The waste sector accounts for ~20% of global anthropogenic methane[2]. In the U.S., landfills remain the third-largest methane source (~14% of national total)[4], with food waste driving ~58% of these fugitive emissions[3].

Yet many high-income economies remain structurally "locked in" to landfill-centric systems. In the U.S., landfilling remains the dominant MSW pathway: in 2018, about 32% of MSW was recycled or composted, 12% was combusted with energy recovery, and roughly 50% was landfilled[5]. Landfilling shares have dropped sharply since 1960 but plateaued in the past decade, supported by low tipping fees, fragmented local governance, and sustained opposition to new combustion facilities. This setting has substantial climate implications, as MSW landfills alone emit methane on the order of tens of millions of passenger vehicles' annual $CO_2$-equivalent[4].

China has followed a very different trajectory. Since the early 2000s, strong central directives to "reduce landfilling and increase harmless treatment" have driven a rapid build-out of waste-to-energy (WtE) incineration capacity[6]. By the early 2020s, MSW incineration had become the dominant urban treatment route, and national daily WtE capacity rose above 1.1 million tonnes[6,7]. This overcapacity has pushed operators to seek additional feedstock, effectively incorporating excavated legacy landfills into broader programs of stock-pollution clean-up and urban land reclamation[8].

These contrasted pathways raise a central puzzle: why have two large economies with comparable access to combustion and emission-control technologies diverged so sharply? Existing work highlights differences in governance and finance[6–8], but there is less systematic analysis of how carbon-accounting rules interact with domestic MSW policy to shape long-run system configurations. This question is critical as other urbanizing countries risk locking in similar high-emission infrastructure.

The rapid rise of artificial intelligence (AI) and cloud computing transforms this domain into a complex System-of-Systems (SoS) challenge. Here, waste management infrastructure (legacy landfills), energy distribution networks (local grids), and digital infrastructure (data centers) function as distinct constituent systems. Global data center electricity consumption is projected to double to ~945 TWh by 2030 [9], with loads concentrated in specific "AI corridors" that strain local grids[10,11]. In the U.S., DCs already consume ~183 TWh (4% of total) and are expected to reach ~426 TWh by 2030 [10]. Evaluating mitigation options in these corridors benefits from cross-layer metrics that connect compute demand, facility efficiency, and system cost [29]. While rising demand suggests a synergy for WtE, the U.S. context—constrained by low tipping fees, the Inflation Reduction Act's focus on renewables, and NIMBY concerns—resists a straightforward "copy-and-paste" of China's centralized model[4,11].

On the surface, rising AI-driven electricity demand and the climate imperative to mitigate landfill methane suggest a natural synergy: advanced WtE and landfill-gas systems could both remediate legacy waste and supply electricity to nearby DCs. China's recent experience, where national WtE capacity now exceeds urban waste volumes, appears to create an exportable pool of technology and engineering expertise [6,7]. Yet in the U.S. context, several constraints make a straightforward "copy-and-paste" expansion of WtE unlikely. High-capex WtE projects must compete with low landfill tipping fees; federal incentives under the Inflation Reduction Act strongly favor wind, solar, storage, and nuclear over incineration; foreign ownership of critical infrastructure is subject to stringent national-security review; and long-standing environmental-justice and "not-in-my-back-yard" (NIMBY) concerns make siting new combustion facilities politically fraught [4,11].

Fig. 1. Conceptual MSW–energy–DC system with legacy landfill remediation

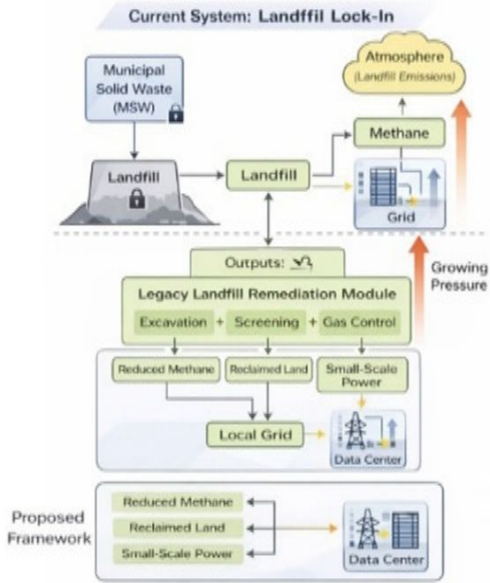

These tensions motivate a shift in perspective. Rather than viewing Chinese-style expansion as a direct template, we argue its value lies in revealing alternative intervention points for system integration. China's experience demonstrates how policy can overcome lock-in and how excavating legacy dumps can simultaneously abate methane and recover urban land[6,8]. Viewed through a systems-engineering lens, these insights suggest that remediation-oriented WtE modules offer a realistic architecture for coupling environmental remediation with energy resilience in landfill-locked jurisdictions.

We therefore ask: under what institutional and carbon-accounting conditions do countries become locked into landfill-centric waste systems, and how can remediation-oriented interventions be designed and deployed to address methane and land constraints while supporting emerging AI and DC loads? We address this overarching question in four steps:

1. (Q1) Develop a comparative systems narrative of MSW management in China and the U.S., identifying policy and financing choices that produced an "incineration transition" in the former and "landfill lock-in" in the latter[5]–[8].

2. (Q2) Construct a stylized analytical framework showing how different carbon-accounting rules and methane valuations translate into distinct socially optimal allocations among landfilling, WtE, and composting.

3. (Q3) Propose a modular waste remediation framework that treats methane abatement and land recovery as primary objectives, with energy recovery as a secondary co-benefit.

4. (Q4) Use a generic AI-corridor scenario to examine how such modules function as complementary resilience resources. Then evaluate the scalability and replicability of this modular SoS architecture, positing it as a generalized template for other regions facing the dual constraints of infrastructure lock-in and rapid digital expansion.

## II. DIVERGENT SYSTEM PATHWAYS: CHINA VS. UNITED STATES

Over the past two decades China and the United States have moved from superficially similar, landfill-dominated waste systems to almost opposite MSW–energy configurations. Using harmonized national statistics, Figure 3 documents the quantitative divergence in disposal structure, WtE use and incineration capacity from 2005–2023, while Figure 2 abstracts these trends into two stylized system architectures. Together they show that the contrast is not a marginal difference in technology mix, but the outcome of distinct governance logics and incentive structures.

### A. Landfill lock-in versus deliberate "de-landfilling"

The starting point in the mid-2000s was broadly comparable: both countries relied predominantly on landfilling, with recycling and composting growing but still secondary. Since then, the trajectories have separated sharply. As illustrated in Figure 3(a), the United States has remained close to an 80% landfill share among disposed MSW, with only modest fluctuations. China, in contrast, has reduced the landfilled share from near universality to below one-fifth by 2023, with the two curves crossing around 2018–2019.

Fig. 2. Contrasting MSW–energy–policy systems in China vs United States

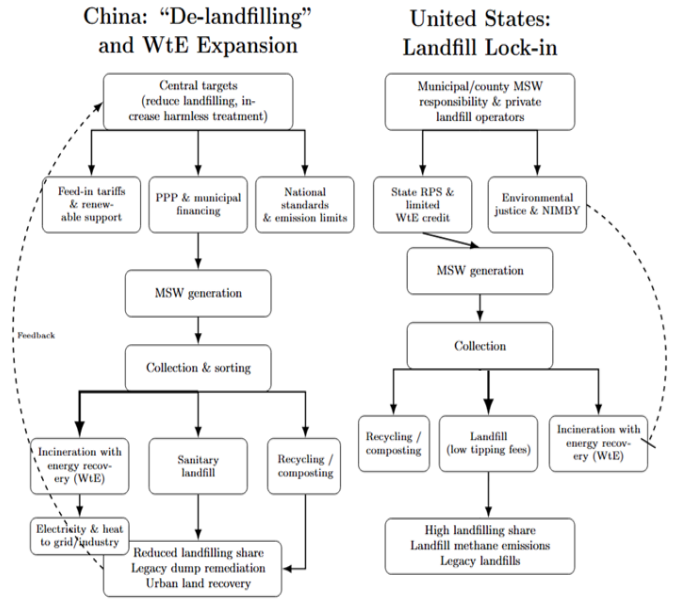

Fig. 3. Diverging waste-to-energy pathways: China vs U.S., 2005–2023

(a) Landfilled share of disposed MSW

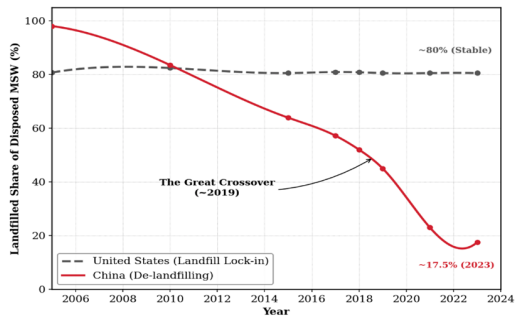

(b) WtE treatment intensity (kg per capita per year)

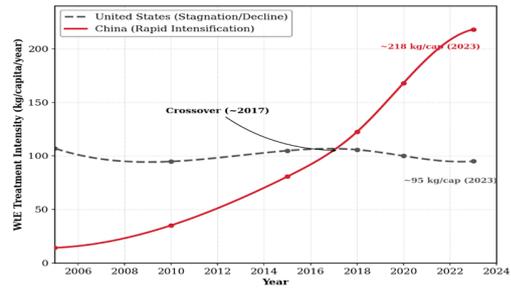

(c) Installed WtE capacity per capita (tonnes/day per 1,000 people)

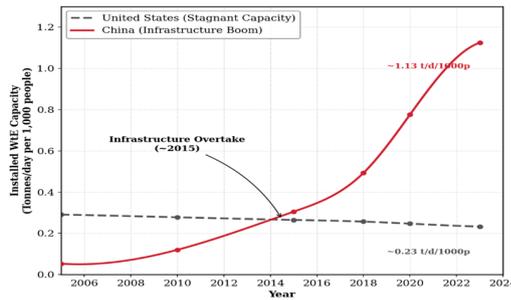

Figure 2 interprets this break as the consequence of different institutional designs. In China, central directives to "reduce landfilling and increase harmless treatment" cascade down through binding targets, performance assessments for local officials, technical standards and emission limits, feed-in tariffs and renewable-energy support, and PPP-based municipal financing. These levers jointly reallocate MSW away from uncontrolled dumps and simple landfills toward sanitary landfills, WtE incineration, and recycling/composting. By contrast, in the U.S. responsibility for MSW sits with municipalities and private landfill operators; there is no national de-landfilling mandate, and low tipping fees make landfilling the budget-minimizing option for local authorities. Strong environmental-justice and NIMBY movements constrain new combustion facilities, but without complementary instruments to raise the shadow price of landfilling, they mainly prevent greenfield WtE rather than accelerate landfill phase-down. The result is a robust landfill-centric equilibrium.

*B. WtE intensity, capacity and system plumbing*

Panels 3(b) and 3(c) show how these institutional differences translate into infrastructure. U.S. WtE treatment intensity has hovered around 90–110 kg per capita per year, and installed capacity per capita has remained near 0.2–0.25 tonnes/day per 1,000 people, consistent with a mature fleet that is largely in maintenance mode. China exhibits a classic build-out pattern: per-capita WtE treatment overtakes the U.S. around 2016–2017 and reaches roughly double U.S. levels by 2023, while installed capacity per capita surpasses the U.S. around 2014–2015 and climbs to several times U.S. levels in the early 2020s.

In system terms (Figure 2), this implies a re-plumbing of the Chinese MSW flow. After collection and sorting, the dominant branch leads to "incineration with energy recovery (WtE)", with sanitary landfills and recycling/composting as complementary outlets. On the U.S. side, the dominant branch remains "landfill (low tipping fees)", with WtE and recycling/composting operating at the margin. China's transition is therefore not simply "more incinerators", but a coordinated reshaping of treatment priorities around WtE and controlled landfilling, backed by central policy and finance. The U.S. system, in contrast, preserves an inherited asset base and price structure that keep landfilling the default.

*C. Legacy landfills as the main fault line*

A final difference concerns how each system treats historical waste stocks. On the Chinese side of Figure 2, the lower node links reduced landfilling with "legacy dump remediation" and "urban land recovery". As WtE and sanitary-landfill capacity expand and land values rise, excavating and remediating old dumpsites becomes administratively and fiscally justifiable, and recovered waste can in some cases be co-processed in WtE plants. On the U.S. side, the corresponding node is "high landfilling share / landfill methane emissions / legacy landfills": historical landfill cells remain large, long-lived methane sources and major land-use constraints, with remediation occurring only in selected Superfund or redevelopment projects.

To summarize, China's pathway is characterized by a centrally coordinated push to de-landfill, supported by tariff and financing instruments that privilege WtE and create space for legacy-site remediation. The United States, by contrast, couples fragmented responsibility and low landfill prices with strong but narrow opposition to new incineration, producing a stable landfill-lock-in in which WtE and remediation play peripheral roles. This institutional divergence sets the stage for the remainder of the paper: the next section formalizes how different carbon-accounting regimes map these configurations into distinct incentive structures for landfilling, current-waste WtE, and legacy-landfill remediation.

III. CARBON ACCOUNTING REGIMES AND SYSTEM-LEVEL INCENTIVES

Recap the second research question, in particular, we are interested in whether legacy landfill remediation with controlled WtE can emerge as a credible mitigation and resilience option under realistic U.S. policies. To answer it, we first develop a compact baseline carbon model for three stylized pathways. We then apply alternative accounting filters by international IPCC conventions, current U.S. practice, and China's de-landfilling regime. Comparing the resulting incentive profiles clarifies why physically similar projects occupy very different positions in each country's mitigation portfolio.

*A. Baseline carbon model for MSW treatment pathways*

Upstream production of goods is excluded; we focus on the incremental effect of alternative treatment options. [1] We consider three generic pathways:

1. Landfill with partial gas capture (current main path)
2. Current-waste WtE incineration + electricity + heat export
3. Legacy landfill remediation + WtE (excavated waste screened, combustible fractions used for energy recovery, and residuals re-landfilled under improved controls).

For each pathway $j$, the net GHG balance per tonne is written schematically as

$$E_j = E_j^{\text{direct}} - B_j^{\text{energy}} - B_j^{\text{methane}}, \quad (1)$$

where $E_j^{\text{direct}}$ aggregates on-site emissions of $CO_2$, $CH_4$, and $N_2O$; $B_j^{\text{energy}}$ denotes avoided grid emissions due to exported electricity or heat; and $B_j^{\text{methane}}$ captures the (discounted) benefit of reducing future landfill methane where applicable.

For landfill with gas capture, we follow the IPCC first-order decay formulation for solid-waste disposal sites[12,14]. For a tonne of MSW with degradable organic carbon (DOC) content $DOC$, methane generation over a 100-year horizon is about

$$M_{\text{CH}_4} = f_{\text{DOC}} \cdot DOC \cdot MCF \cdot F, \quad (2)$$

where $f_{\text{DOC}}$ is the fraction of DOC that decomposes, $MCF$ is the methane correction factor, and $F$ is the fraction of $CH_4$ in landfill gas. Capture efficiency $R$ and oxidation factor $OX$ then yield net methane emissions[2]

$$M_{\text{CH}_4}^{\text{net}} = M_{\text{CH}_4}(1-R)(1-OX). \quad (3)$$

For current-waste WtE, we treat fossil and biogenic carbon separately in line with IPCC incineration guidelines[13]. Direct fossil $CO_2$ emissions per tonne are

$$E_{\text{WtE}}^{\text{fossil}} = C^{\text{fossil}} \cdot \frac{44}{12}, \quad (4)$$

where $C^{\text{fossil}}$ is the fossil carbon content per tonne MSW. Biogenic $CO_2$ is conventionally treated as climate-neutral in inventory accounting, while $N_2O$ and upstream auxiliary fuel use are included explicitly[13,16-17]. Avoided grid emissions are computed from net electricity export and a marginal emission factor representing the displaced generation mix [15,18]. Recent LCA studies suggest that modern WtE facilities achieve net climate impacts of roughly 0.3–0.5 t $CO_2$e per tonne MSW, often 60–80 % lower than comparable landfilling under similar grid and capture assumptions [16-18].

The third pathway, legacy landfill remediation + WtE, operates on already-disposed waste. Excavation and screening remove a portion of the remaining degradable organics and recover combustible fractions that can be co-fired in WtE or cement kilns[19-20]. Residuals are re-landfilled with improved covers, gas capture, or biocovers[21]. In the model, this pathway reduces future methane by lowering the effective DOC in situ and by improving $R$ and $OX$ for the remaining mass. It adds one-off emissions from excavation and transport, plus any fossil $CO_2$ from energy recovery, but also yields avoided grid emissions analogous to pathway (2). The growing landfill-mining literature reports potential methane reductions of 20–30 % or more at remediated sites[19-21].

In summary, the baseline model compresses the main physical components, methane generation and capture, fossil versus biogenic carbon, and energy credits, into three comparable expressions $E_j$. These expressions can then be re-weighted by different accounting regimes without altering the underlying physics, allowing us to ask how the same physical project looks to national inventories, local regulators, and corporate buyers in China and the United States.

*B. Carbon accounting regimes and incentives: U.S. vs. China*

*International baseline (IPCC conventions).* Both China and the U.S. compile inventories under IPCC guidelines[12-14], by which landfills are reported in the waste sector, using Equations (2)–(3). [3] The IPCC baseline thus recognizes the physical superiority of high-capture landfills and well-designed WtE over uncontrolled disposal, but it does not by itself determine private incentives. That depends on how each country translates inventory metrics into taxes, subsidies, and appraisal rules.

*United States: landfill lock-in and WtE ambiguity.* In the U.S., inventory accounting is overlaid with a patchwork of policies and market instruments[15-18]. First, federal and state programs grant renewable-energy credits and tax incentives to landfill-gas-to-energy projects, treating captured methane as a renewable resource. Second, most WtE plants are regulated primarily as combustion sources under air-pollution rules; while some qualify for state-level renewable credits, they are largely excluded from the strongest federal incentives (investment and production tax credits under the Inflation Reduction Act[17]). Third, legacy landfill remediation is typically funded through site-specific environmental or brownfield programs, with climate benefits either ignored or treated as co-benefits[19,21].

Mapped onto our baseline model, these choices imply that:

- Pathway (1) landfill with gas capture has multiple revenue streams (tipping fees, electricity sales, environmental credits), making its private net cost often lower than suggested by its physical GHG impact.

- Pathway (2) current-waste WtE faces higher capital and regulatory hurdles, and its climate performance is sometimes viewed as ambiguous because biogenic $CO_2$ is discounted while stack emissions remain visible [16-18].

---

[1] To compare regimes consistently, we adopt a common functional unit: the net 100-year global-warming impact (GWP$_{100}$, in t $CO_2$e) per tonne of waste managed, either current MSW or excavated legacy waste.

[2] We convert methane to $CO_2$e using IPCC GWP$_{100}$ values and subtract any electricity produced from collected gas using standard emission factors for displaced grid power [12], [15]. Parameter ranges are calibrated from IPCC guidelines and the U.S. GHG inventory [12], [14], [15], giving typical net impacts on the order of 1.5–2.0 t $CO_2$e per tonne MSW for landfills with partial gas utilization, and higher values where capture is limited.

[3] WtE stack emissions are reported in the energy sector, with biogenic $CO_2$ recorded as "memo items"; and legacy landfill mining appears only indirectly through adjustments to waste quantities or oxidation assumptions. Methane carries a high GWP but is not explicitly priced, and electricity credits are reflected only if countries choose to allocate them in national mitigation planning, not in the inventory itself.

- Pathway (3) legacy remediation + WtE lacks a stable business model: excavation and site remediation costs are high with no mechanism to monetize avoided methane.

As a result, a tonne of waste physically better managed through WtE or remediation may still appear financially inferior to continued landfilling with incremental gas capture.

*China: de-landfilling + WtE as environmental infrastructure.* China applies the same IPCC inventory framework but embeds MSW policy within a different institutional logic. Central targets to reduce landfilling and increase harmless treatment are tied to local performance evaluation; WtE plants are treated as environmental infrastructure, benefiting from feed-in tariffs or guaranteed off-take, and landfill mining is more integrated into urban land-reclamation and pollution-control [6,7,16,19,20]. In terms of our model, this regime:

- Implicitly assigns a high shadow price to methane and urban land: advocate projects reducing landfill $CH_4$/free up land.

- Ensures that WtE plants recover capital through long-term power-purchase agreements and tipping fees.

- Encourages legacy remediation as part of broader land-use and environmental campaigns[19,20].

Thus, the same physical pathways occupy very different positions in the Chinese mitigation portfolio: WtE and landfill mining appear as central tools for achieving mandatory targets, rather than marginal competitors to cheap landfilling.

*C. Stylized incentive profiles and synthesis*

To synthesize these findings, we construct a one-tonne incentive map, shown in Table 1. For each pathway, we identify which components of Equation (1) are monetized or rewarded, which are penalized, and which remain invisible to decision-makers. For U.S. landfilling with gas capture often has positive net incentives: tipping fees and LFG-to-energy credits more than offset the cost of residual methane[15],22. Current-waste WtE yields substantial physical mitigation relative to landfilling[16-18], but once future biogenic $CO_2$ is discounted and strong federal incentives are reserved for wind, solar, storage, and nuclear, its private climate premium is weak or even negative. Legacy remediation plus WtE offers large potential reductions in site-level methane and land-use risks [19]–[21], yet these benefits are only partially recognized, leaving the pathway socially efficient in some locations but privately unattractive.

The Chinese regime reorders this incentive map. Central performance targets and financing tools elevate pathways that reduce landfill dependence and methane emissions; WtE plants are treated as priority environmental infrastructure; landfill-mining projects can draw on renewal and pollution-control budgets. In effect, more of the physically important terms in Equation (1), especially methane abatement and land recovery, are aligned with local officials' and project developers' objective functions than in the U.S.

TABLE I. TABLE TYPE STYLES

| Pathway | U.S. regime: what is rewarded? | U.S.: what is ignored/penalized? | China regime: what is rewarded? | China: what is ignored/penalized? |
|---|---|---|---|---|
| (1) Landfill + gas capture | Tipping fees; LFG power as "renewable" in RPS | Residual $CH_4$ over 100 years | Some landfill upgrades where needed | Long-term $CH_4$ still costly for targets |
| (2) Current-waste WtE | Sometimes tipping fee + some power revenue | Often excluded from strongest clean-energy credits | Tariffs, PPPs, environmental-infrastructure framing | Local air-quality constraints |
| (3) Legacy remediation + WtE | Brownfield grants (site-specific); limited climate credit | Most avoided $CH_4$ and land benefits invisible | Integrated into "stock pollution" + land-reclamation plans | Energy is still a secondary co-benefit |

To summarize, under the same physical technology set, different accounting and incentive regimes produce different rankings of the three pathways. In a U.S.-style regime, incremental gas capture at existing landfills dominates, while WtE and remediation struggle to compete. In a Chinese-style regime, de-landfilling and legacy remediation move to the center of the mitigation strategy. From a systems perspective, this suggests that any realistic role for legacy landfill remediation in U.S. decarbonization and DC resilience will depend less on discovering new technologies than on adjusting the accounting and incentive filters applied to the already-available pathways.

A qualitative sensitivity analysis reveals that the system hierarchy is highly responsive to the shadow price of methane. While landfilling dominates the objective function under low-carbon-price regimes, our model implies a structural tipping point: as $\lambda$ rises to reflect the 20-year Global Warming Potential of methane (GWP20 ≈ 80) rather than the standard 100-year metric, the avoided liability from future emissions drives the net social cost of Pathway 3 (Remediation) below that of the 'monitor-and-maintain' baseline. This suggests that the lock-in is not technically immutable but parametrically sensitive to specific valuation coefficients within the accounting filter.

Environmental Trade-off Validation. A common critique of WtE expansion concerns direct combustion emissions. However, a systems-level mass balance reveals that the 'methane multiplier' dominates the net climate impact. Since fugitive methane possesses a Global Warming Potential (GWP) 28-84 times that of $CO_2$ (depending on the time horizon), the abatement achieved by permanently capping this persistent source decisively outweighs the direct fossil-$CO_2$ penalty of incineration. Furthermore, the proposed modular units incorporate advanced 7-step flue gas treatment technologies verified to meet or exceed stringent EU Industrial Emissions Directive (2010/75/EU) standards. Thus, rather than increasing pollution, the remediation pathway functions as a net-negative carbon intervention that simultaneously resolves local toxicity risks.

IV. A LEGACY WASTE REMEDIATION FRAMEWORK INFORMED BY CHINESE EXPERIENCE

Section 3 showed that under many carbon-accounting regimes, the socially preferred path is to curb landfill methane and reclaim constrained land, while treating energy recovery as a secondary co-benefit. In this section we translate that insight into an engineering-style remediation framework, drawing on Chinese practice and mapping it to U.S. legacy landfills located near emerging AI "corridors".

## A. Design principles and Chinese technical modules

Recent Chinese projects have moved beyond simple capping and gas flaring toward integrated "landfill mining + remediation" schemes that excavate aged waste, separate fractions, tighten gas control, and prepare land for redevelopment [23,24]. Life-cycle assessments suggest that well-designed projects can cut long-run methane emissions from targeted sites by roughly 60–70 %, while converting part of the combustible fraction and landfill gas into useful energy and re-using soil-like material on site[25].

Fig. 4. Modular legacy landfill remediation system (energy and DC loads)

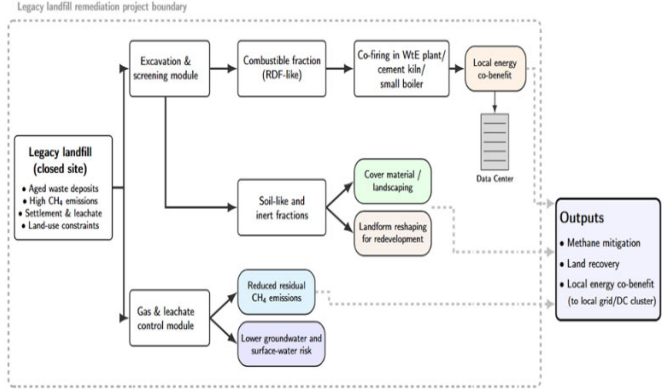

Figure 4 summarizes the modular system architecture we adopt. A legacy capped landfill is treated as a stock of embedded risks and resources: aged deposits, high $CH_4$ emissions, settlement and leachate issues, and land-use constraints. Within a defined project boundary, two main technical modules operate:

1. Excavation and screening module. Aged waste is excavated and mechanically separated into: (i) a combustible, refuse-derived fuel (RDF-like) fraction that can be co-fired in existing WtE plants, cement kilns, or modular small boilers; (ii) soil-like and inert fractions used as engineered cover or for landform reshaping and landscaping.

2. Gas and leachate control module. Remediation is combined with upgraded gas wells, biocovers, and leachate collection/treatment, reducing residual $CH_4$ emissions and groundwater risk [26,27].

The outputs are threefold: (i) methane mitigation, (ii) land recovery suitable for controlled redevelopment, and (iii) local energy co-benefits that can be connected to a nearby microgrid or DC cluster. This framing treats WtE-type technologies not as stand-alone power plants, but as modular environmental-remediation tools with energy as a by-product.

## B. Existing practice versus remediation-based option

Table 2 contrasts a conventional "monitor and maintain" strategy for a capped landfill with a time-limited modular excavation and remediation campaign. Under current practice, most closed landfills remain long-term environmental liabilities: methane continues to be generated with only partial capture, settlement and leachate risks persist, and the footprint is effectively frozen as low-value open space. Small landfill-gas-to-energy units, where present, typically provide modest and declining generation tied to the bulk grid[27].

By contrast, the remediation-based option reframes the site as a finite clean-up project (≈18–36 months of active work within a multi-year program). Excavation and enhanced gas control can deliver a large reduction in long-run $CH_4$ emissions, while part of the footprint is engineered for reuse, potentially as a platform for co-located DC or grid assets with green buffers. Combustible fractions and upgraded landfill-gas collection enable a 10–30 MW class combined heat-and-power (CHP) unit, designed to operate behind the meter within a local microgrid. At the same time, the project creates visible local benefits, risk reduction, land recovery, and construction/remediation jobs, which can be framed as environmental-justice improvements rather than a new incinerator siting dispute[23,24].

In short, Table 2 makes explicit that remediation-based projects shift the emphasis from "minimizing ongoing compliance cost" to "monetizing risk reduction and resilience gains", including for nearby AI and cloud-computing loads.

## C. A generic "AI-corridor landfill" scenario

To make these trade-offs more concrete, we construct a stylized scenario for a legacy landfill located near an emerging AI DC corridor (Table 3). The numbers are indicative rather than site-specific, calibrated to typical EPA and IPCC factors for U.S. landfills and to representative WtE/CHP performance [23,25,27].

TABLE II.  TABLE TYPE STYLES

| Aspect | Baseline capped landfill | Modular excavation & remediation |
|---|---|---|
| Project framing & timeline | Long-term monitoring asset; incremental well/cap upgrades; limited new capacity; permitting under waste/air rules. | Time-limited remediation campaign (~18–36 months); framed as environmental clean-up; modular, containerized units can be redeployed. |
| Methane & environmental risk | Continuing CH4 generation with partial capture; residual fugitive emissions for decades; persistent leachate/settlement risk. | Excavation plus enhanced gas control and biocovers; large reduction in long-run CH4; targeted reduction of leachate and geotechnical risk. |
| Land use & planning | Site effectively frozen; mainly low-value open space; constrained redevelopment near AI corridors. | Portion of footprint engineered for reuse; potential platforms for co-located DC or grid assets plus green buffers. |
| Power & resilience value | Small, often declining LFG-to-energy units tied to the bulk grid; limited role in stress events. | 10–30 MW firm CHP from combustible fractions and LFG; behind-the-meter microgrid option; supports cooling load and black-start/islanding. |
| Asset & community impacts | Compliance cost with limited local benefits; long-run EJ concerns from odour, CH4, and groundwater risk. | Visible risk reduction; land recovery; remediation local jobs; EJ safeguards, easier than new mass-burn plants. |

TABLE III.  TABLE TYPE STYLES

| Metric | Baseline capped landfill | Modular remediation project (10-year campaign) |
|---|---|---|
| Annual CH4 emissions (tCO2e/yr) | ~200,000 | ~60,000–80,000 (≈ 60–70 % ↓) |
| Land available for redevelopment | ≈0 ha | ~25 ha of engineered platform |
| Firm electrical capacity | 0 MW (or small, declining LFG unit) | ~20 MW CHP (combustible fraction + LFG) |
| Annual electricity output | <30 GWh | ~150 GWh |

We consider a closed landfill with several tens of millions of tonnes of aged waste and current methane emissions on the order of 200,000 $tCO_2e$ per year, consistent with Tier-1 inventory estimates for large U.S. sites[27]. Under a remediation campaign that combines excavation, enhanced gas capture, and engineered biocovers, landfill-mining studies suggest that 60–70 % reductions in long-run $CH_4$ emissions are technically achievable for the treated portion of the site [23,24]. This yields a residual of roughly 60,000–80,000 $tCO_2e$ per year in Table 3.

Excavation and re-grading are assumed to free up around 25 ha of engineered platform, a footprint compatible with either a medium-size DC, an on-site substation plus battery storage, or other grid-supporting assets. Combustible fractions extracted during mining, together with improved landfill-gas capture, are routed to a ≈20 MW CHP unit operating at baseload with a high capacity factor (≈0.85–0.9), reflecting the near-continuous fuel availability of MSW-based systems[25]. This implies annual electricity generation of roughly 150 GWh, which Table 3 benchmarks against the load of a representative 300 MW hyperscale DC (≈2.5–2.6 TWh/yr at high utilization).

In pure energy terms, the remediation project supplies only about 5–6 % of the DC's annual demand, confirming that WtE cannot "solve" the AI electricity gap on its own. However, the value of this power lies in its firm, behind-the-meter character. By directly feeding a local microgrid or critical cooling loads, the 20 MW CHP plant can (i) relieve pressure on congested transmission nodes, (ii) displace part of the diesel backup capacity, and (iii) provide limited black-start and ride-through capability during grid disturbances. These resilience and risk-reduction benefits are incremental to the primary objectives of methane abatement and land reclamation.

Taken together, Figure 4 and Tables 2-3 operationalize our third research question, our framework suggests that the answer is yes, provided projects are framed as environmental remediation rather than greenfield power plants, rely on modular and re-deployable equipment, and are integrated into local microgrid and land-use planning rather than bulk-grid baseload planning.

*D. AI DC Corridors, Resource Planning, and Implications*

Q4 focuses how remediation-oriented WtE modules, once defined and sized as in Section IV-C, could play a system-relevant role in AI DC corridors (often multi-campus clusters with corridor-level constraints) under tight deployment and resilience constraints[30]. Hyperscale AI DCs in the United States are emerging in a handful of "AI corridors" where load growth is steep, transmission interfaces are congested, and interconnection queues for new large generators or grid upgrades often span many years[9-11]. Within a five-year horizon, most operators therefore rely on a mix of natural-gas peaker plants, battery systems (e.g., emerging Zn-based chemistries) [31], and expanded diesel backup, while longer-lead options such as new nuclear units or major transmission lines remain largely out of reach.

Against this backdrop, a remediation-based landfill project is not a competitor to wind, solar, or nuclear for bulk energy, but a niche resource with distinctive attributes. As Section IV-C showed, a single large-site project can realistically supply only about 5–6 % of a 300 MW hyperscale DC's annual load, yet it can do so as tens-of-megawatts of firm capacity that is physically adjacent to the DC and deployable within roughly 18–36 months when delivered via modular, containerized excavation and CHP units under an environmental-remediation permitting pathway [23,27,28]. In resource-planning terms, the key question is therefore not "can remediation power the corridor?", but "what is the value of a small, colocated, high-capacity-factor module in a corridor that is both grid-constrained and politically sensitive?"

The system value concentrates along three dimensions. First, co-locating the remediated landfill platform and its 10–30 MW CHP unit with a DC cluster enables a behind-the-meter microgrid architecture.[4] This reduces dependence on uncertain bulk-grid upgrades, alleviates flows across already saturated transmission nodes, and allows part of the DC's cooling and auxiliary loads to be served from local firm generation rather than regional peakers. Second, because the CHP unit operates at high utilization and can be sized around critical thermal and IT loads, it provides a 24/7 firm block that partially substitutes for battery storage and diesel backup, lowering outage risk and improving effective reliability without requiring large-scale storage build-out. Quantitatively, while serving only ~5-6% of peak demand, this islandable capacity is dimensioned to sustain 100% of critical thermal management and network orchestration loads, effectively transforming a catastrophic Loss of Load event into a survivable partial curtailment state. Third, during extreme events, winter storms, heatwaves, or hurricane-related outages, the combination of on-site fuel stock (aged waste plus landfill gas) and islanding capability can support black-start and limited island operation, extending ride-through time beyond what is feasible with conventional diesel tanks alone.

These characteristics have planning and policy implications that differ from those of conventional WtE plants. Rather than seeking generic feed-in tariffs or competing head-to-head with renewables in capacity auctions, remediation-based modules are more naturally framed as methane-mitigation and brownfield-reuse projects with co-produced resilience services. For regulators and system operators, this suggests small pilot programs in suitable AI corridors where legacy landfills already constrain urban form and contribute materially to $CH_4$ inventories, with explicit treatment of remediated sites and associated CHP units as resilience assets in local resource plans. For DC owners, it points to a new class of partnerships, structured around remediation contracts, brownfield land leases, or methane-mitigation crediting, through which AI infrastructure can co-finance projects that simultaneously reduce climate risk, expand developable land, and strengthen local power-system resilience. In this sense, Q4 is answered not by proposing WtE as a bulk-power solution for AI, but by positioning remediation-based modules as targeted, time-limited components of a diversified decarbonization and resilience portfolio for AI DC corridors..

V. CONCLUSIONS

This paper has examined why China and the United States, two large economies with access to similar combustion and control technologies, have diverged into "incineration transition" and "landfill lock-in" regimes, and what this divergence implies for AI-era electricity planning. Using harmonized national statistics, we showed that China has driven the landfilled share of MSW below one-fifth while building per-capita WtE capacity more than four times U.S. levels, whereas the United States has remained in a stable, landfill-dominant configuration. A simple carbon-accounting framework then demonstrated how different methane valuations and inventory rules shift the socially preferred mix of landfilling, WtE, and composting, and how these choices interact with corporate decarbonization strategies for data centers.

---

[4] Technically, this interface utilizes a fast-acting Point of Common Coupling (PCC) switchgear managed by a microgrid controller. The CHP genset must be equipped with isochronous governance capability to transition from grid-following (PQ control) to grid-forming (V-f control) mode within milliseconds during a disturbance, ensuring voltage stability and seamless islanding for sensitive data center loads.

Building on recent Chinese practice, we proposed a modular legacy landfill remediation system that treats methane abatement and land recovery as primary objectives and energy recovery as a co-benefit. The stylized AI-corridor scenario suggests that a single project at a large legacy landfill can plausibly reduce long-run methane emissions by 60–70 %, reclaim tens of hectares of developable land, and provide on the order of 150 GWh/yr of firm, behind-the-meter electricity, roughly 5–6 % of a 300 MW hyperscale data center's demand. This contribution is modest in energy terms but system-salient: it improves local reliability, reduces reliance on diesel backup, and repurposes an environmental liability into a brownfield–data-center asset.

For planners and regulators, the main implication is that remediation-oriented WtE modules should not be evaluated as competing bulk-power resources, but as targeted, time-limited interventions that jointly address methane, land scarcity, and AI-driven resilience needs in specific corridors. Future work could embed these modules in formal capacity-expansion and microgrid-planning models, test alternative policy designs for methane crediting and brownfield leasing, and extend the comparative analysis to other rapidly urbanizing regions deciding between landfill expansion, WtE build-out, and remediation-first strategies.

ACKNOWLEDGMENT